\documentclass[a4paper,11pt]{article}
\usepackage{pos}

\usepackage{amsmath}
\usepackage{sidecap}

\newcommand{\be}{\begin{equation}}
\newcommand{\ee}{\end{equation}}
\newcommand{\bea}{\begin{eqnarray}}
\newcommand{\eea}{\end{eqnarray}}

\title{Window contributions to the muon hadronic vacuum polarization with twisted-mass fermions}
\ShortTitle{Window contributions to $a_\mu^{\rm HVP}$ with TM fermions}

\author*[a]{D.~Giusti}
\author[b]{S.~Simula}

\affiliation[a]{Universit\"at Regensburg, Fakult\"at f\"ur Physik,\\
  D-93040, Regensburg, Germany}

\affiliation[b]{Istituto Nazionale di Fisica Nucleare, Sezione di Roma Tre,\\
  Via della Vasca Navale 84, I-00146, Rome, Italy}

\emailAdd{davide.giusti@ur.de}

\abstract{We present a lattice calculation of the Euclidean position-space windows contributing to the leading-order hadronic vacuum polarization term of the muon anomalous magnetic moment $a_\mu$.
Short-, intermediate- and long-distance windows are considered in order to isolate different scales sensitive to specific integration ranges of experimental time-like data used in the R-ratio.
By adopting the same smooth window function introduced by the RBC and UKQCD Collaborations with width parameter $\Delta = 0.15~\rm fm$, for the isospin-symmetric, light, quark-connected component we get $a_\mu^{\rm SD} (ud) = 48.21\,(80) \cdot 10^{-10}$ , $a_\mu^{\rm W} (ud) = 202.2\,(2.6) \cdot 10^{-10}$ and $a_\mu^{\rm LD} (ud) = 382.5\,(11.7) \cdot 10^{-10}$ in the short- (SD), intermediate- (W) and long-distance (LD) time regions, respectively, with $t_0 = 0.4~\rm fm$ and $t_1 = 1.0~\rm fm$.
Our results are obtained using the gauge configurations generated by the Extended Twisted Mass Collaboration with $N_f=2+1+1$ dynamical quarks, at three values of the lattice spacing varying from 0.089 to 0.062 fm, at several lattice volumes and with pion masses in the range $M_\pi \simeq 220 - 490~\rm MeV$.}

\FullConference{%
 The 38th International Symposium on Lattice Field Theory, LATTICE2021
  26th-30th July, 2021
  Zoom/Gather@Massachusetts Institute of Technology
}

\begin{document}
\maketitle

\section{Introduction}
\label{sec:intro}

The muon magnetic moment anomaly, $a_\mu = (g - 2)/2$, exhibits a long-standing discrepancy between the Standard Model (SM) prediction and the experimentally measured value.
Since this tension, if confirmed with high significance, might provide an indirect evidence for new physics beyond the SM, an intense research program is currently underway in order to achieve a significant reduction of the experimental and theoretical uncertainties.

The new Fermilab Muon $g-2$ (E989) experiment has recently presented its first results for the positive muon magnetic anomaly, analyzing Run-1 measurements collected during the data taking in 2018.
The value of $a_\mu$, determined with an accuracy of 0.46 ppm \cite{Muong-2:2021ojo}, is found to be in excellent agreement with the previous E821 measurement at BNL \cite{Muong-2:2006rrc}, while it differs from the SM prediction by $3.3\sigma$.
Data analyses on the second and third runs of the E989 experiment are underway and, by combining the results from all runs, a final sensitivity four-times better than the E821 determination is expected to be reached.
An alternative low-energy approach at J- PARC is expected to reach a precision similar to the existing BNL measurement.

On the theoretical side the present accuracy of the SM prediction is at 0.53 ppm \cite{Zyla:2020zbs}.
To leverage the new experimental efforts, the theory errors must be reduced to the same level as the experimental uncertainties.
The main contribution to $a_\mu$ comes from quantum electrodynamics (QED) and can be accurately computed using a perturbative expansion in the fine-structure constant $\alpha_{em}$ \cite{Aoyama:2012wk,Aoyama:2019ryr}.
The small electroweak corrections are also under control \cite{Czarnecki:2002nt,Gnendiger:2013pva}.
Finally, although quarks and gluons do not couple directly to the muon, they do interact via loop diagrams.
Even if hadronic contributions are relatively small, they completely dominate the error budget and are the limiting factor in view of reducing the theory error.
The dominant sources of uncertainty in the SM prediction are from two distinct contributions: the hadronic vacuum polarization (HVP) that starts to ${\cal O}(\alpha_{em}^2)$ and the hadronic light-by-light scattering (HLbL) contributions entering at ${\cal O}(\alpha_{em}^3)$.

There are a number of complementary theoretical efforts underway to better understand and quantify these hadronic corrections, including dispersive methods, lattice QCD, and effective field theories, as well as a number of different experimental efforts to provide inputs to dispersive, data-driven evaluations.
A concerted effort of the theory community to improve upon and scrutinize the existing SM results has been made possible thanks to the formation of the Muon $g-2$ Theory Initiative and a Whitepaper summarizing the current theory status has been recently finalized \cite{Aoyama:2020ynm}.
The main outcome is that for $a_\mu^{\rm HVP}$ the overall lattice precision is not yet competitive with respect to the one of the dispersive results, while recent lattice estimates of the HLbL term are consistent with the phenomenological and dispersive findings within the current level of precision and rule out the HLbL contribution as an explanation for the current tension between theory and experiment.

For the HVP contribution, however, tensions exist within lattice QCD calculations as well as between lattice QCD calculations and R-ratio results.
Recently the BMW collaboration \cite{Borsanyi:2020mff} claims to have reached a precision for $a_\mu^{\rm HVP}$ similar to the one of the dispersive approaches, although getting a $2.1\sigma$ discrepancy for the central values.
At this point, the lattice calculations exhibiting a tension with R-ratio results share some aspects.
They are performed at physical pion mass, with staggered sea quarks and use inverse lattice spacings in the range from $a^{-1} \approx 1.6$ GeV to $a^{-1} \approx 3.5$ GeV.
Concretely, there are tensions for the isospin-symmetric quark-connected light-quark contribution, which provides almost $90\%$ of the total $a_\mu^{\rm HVP}$.

Matching the precision of the new experiments requires to determine the HVP contribution at the per-mille level and this represents an enormous challenge for lattice simulations.
Since current estimates for this observable are usually dominated by systematic errors, it is of major importance to perform further cross-checks between collaborations to provide evidence that lattice simulations are under control.
In particular the topical workshop by the Muon $g-2$ Theory Initiative ``The hadronic vacuum polarization from lattice QCD at high precision'' held online in November 2020 (\url{https://indico.cern.ch/event/956699/}) has offered a platform to compare lattice results for Euclidean position-space windows contributing to $a_\mu^{\rm HVP}$.
The advantage of those quantities, defined in the following section, is that, by choosing an appropriate window, the calculation can be made much less challenging on the lattice than for the full $a_\mu^{\rm HVP}$.
Those observables, being less sensitive to some specific source of systematic error, are considered to be particularly well-suited benchmark candidates for comparing different lattice methods.
In this respect, many collaborations have presented their preliminary determinations for the window contributions in numerous talks at this Lattice conference.

In this contribution we present our results for the short-, intermediate- and long-distance windows of the isospin-symmetric, light, quark-connected component of $a_\mu^{\rm HVP}$ using the QCD gauge configurations generated by ETMC with $N_f = 2 + 1 + 1$ dynamical quarks, at three values of the lattice spacing varying from 0.089 to 0.062 fm, at several values of the lattice spatial size $(L \simeq 1.8 \div 3.5 ~ {\rm fm})$ and with pion masses in the range between $\approx 220$ and $\approx 490$ MeV (details concerning the 17 ETMC gauge ensembles can be found in Table 1 of Ref.~\cite{Giusti:2018mdh}).
For further readings on our lattice determinations of $a_\mu^{\rm HVP}$ we refer the interested reader to Refs.~\cite{Giusti:2018mdh,Giusti:2017jof,Giusti:2019xct,Giusti:2019hkz,Giusti:2020efo}.

\section{Definitions}
\label{sec:defs}

The light-quark contribution to the HVP term of the muon anomalous magnetic moment can be calculated by adopting the time-momentum representation \cite{Bernecker:2011gh}
\be
    a_\mu^{\rm HVP} (ud) = 4 \alpha_{em}^2 \frac{1}{m_\mu^2}\int_0^\infty dt ~ K(m_\mu t) V^{ud}(t) ~ ,
    \label{eq:amu_t}
\ee
where the kernel function $K$ is given by \footnote{In Eq.~(\ref{eq:kernel}) $j_0(y)$ is the spherical Bessel function $j_0 = \sin{(y)} / y$.}
\be
    K(z) = z^2 \int_0^1 dx (1-x) \left[ 1 - j_0^2 \left(\frac{z}{2}\frac{x}{\sqrt{1-x}} \right) \right] 
    \label{eq:kernel}
\ee
and $V^{ud}(t)$ is the vector current-current Euclidean correlator defined as
\be
    V^{ud}(t) \equiv -\frac{1}{3} \sum_{i=1,2,3} \int d\vec{x} ~ \langle J^{ud}_i(\vec{x}, t) J^{ud}_i(0) \rangle ~ ,
    \label{eq:VV}
\ee
with $t$ being the Euclidean time distance and
 \be
     J^{ud}_\mu(x) \equiv \sum_{f = u, d} q_f ~ \overline{\psi}_f(x) \gamma_\mu \psi_f(x) ~ .
     \label{eq:Jem}
 \ee

It is instructive to isolate specific ranges of Euclidean time in order to better understand their contributions to $a_\mu^{\rm HVP} (ud)$.
This can be accomplished by constructing windows that suppress contributions outside of the window region.
The window method has been presented for the first time in Ref.~\cite{RBC:2018dos} as a tool to improve the accuracy of the HVP by supplementing the dispersive results based on R-ratio measurements with lattice inputs in a time-region where the lattice data turn out to be more precise.
Rather than using Heaviside step functions to isolate these ranges, which would have significant dependence on the lattice cutoff near the boundary of the window, a smoothed step is considered \cite{RBC:2018dos}
\be
\Theta(t,t^\prime;\Delta) = \frac{1}{1+e^{-2(t-t^\prime)/\Delta}} ~ .
\label{eq:winfunc}
\ee
This step function suppresses all values below $t^\prime$ and has a width parameterized by $\Delta$.
From these step functions, windows into specific regions of $a_\mu^{\rm HVP} (ud)$ Euclidean time can be studied by instead convoluting the integrand of Eq.~(\ref{eq:amu_t}) with the smooth window function (\ref{eq:winfunc}).

In what follows we consider the contributions of three separate windows, namely
\bea
\label{eq:winshort}
a_\mu^{\rm SD}(t_0;\Delta) & \equiv & 4 \alpha_{em}^2 \frac{1}{m_\mu^2}\int_0^\infty dt ~ K(m_\mu t) V^{ud}(t) \left[ 1 - \Theta(t,t_0;\Delta) \right] ~ , \\
\label{eq:winint}
a_\mu^{\rm W}(t_0,t_1;\Delta) & \equiv & 4 \alpha_{em}^2 \frac{1}{m_\mu^2}\int_0^\infty dt ~ K(m_\mu t) V^{ud}(t) \left[ \Theta(t,t_0;\Delta) - \Theta(t,t_1;\Delta)\right] ~ , \\
\label{eq:winlong}
a_\mu^{\rm LD}(t_1;\Delta) & \equiv & 4 \alpha_{em}^2 \frac{1}{m_\mu^2}\int_0^\infty dt ~ K(m_\mu t) V^{ud}(t) \, \Theta(t,t_1;\Delta) ~ ,
\eea
where the function $\Theta(t,t^\prime;\Delta)$ is defined in (\ref{eq:winfunc}) and the parameters $t_0, t_1$ and $\Delta$ are chosen to be
\be
t_0 = 0.4 ~ {\rm fm} ~ , \qquad t_1 = 1.0 ~ {\rm fm} ~ , \qquad \Delta = 0.15 ~ {\rm fm} ~ .
\ee

By design, the choice of the above parameters leads to several advantages for the intermediate window (\ref{eq:winint}), since both the short-distance region, where large cutoff effects are present, and the long-distance region, where the statistical uncertainties and finite-volume effects (FVEs) are large, are cut away.

\section{Effective lepton mass and effective windows}
\label{sec:EW}

To perform the calculation of the three windows defined in (\ref{eq:winshort})-(\ref{eq:winlong}) we generalize the ETMC effective lepton mass procedure introduced in Ref.~\cite{Burger:2013jya}.
Namely, we assume effective values both for the lepton mass $m_\mu^{eff}$ and for the parameters $t_0^{eff}, t_1^{eff}$ and $\Delta^{eff}$ defined as
\bea
\label{eq:mmueff}
m_\mu^{eff} & \equiv & \left( m_\mu / X^{phys} \right) X ~ , \\
\label{eq:t0eff}
t_0^{eff} & \equiv & t_0 X^{phys} / X ~ , \\
\label{eq:t1eff}
t_1^{eff} & \equiv & t_1 X^{phys} / X ~ , \\
\label{eq:Deltaeff}
\Delta^{eff} & \equiv & \Delta \, X^{phys} / X ~ ,
\eea
where $X$ is a hadronic quantity having the dimension of a mass, which can be extracted from lattice correlators, and $X^{phys}$ is its value at the physical point.
In what follows we refer to the choices (\ref{eq:mmueff}-\ref{eq:Deltaeff}) as the effective lepton mass (ELM) and effective window (EW) procedure.

Thus, in the case of the intermediate window (taken as an example) we get
\bea
a_\mu^{\rm W}(t_0^{eff},t_1^{eff};\Delta^{eff}) & \equiv & 4 \alpha_{em}^2 \left(\frac{1}{m_\mu^{eff}}\right)^2\int_0^\infty dt ~ K(m_\mu^{eff} t) V^{ud}(t) \nonumber \\
& \cdot & \left[ \Theta(t,t_0^{eff};\Delta^{eff}) - \Theta(t,t_1^{eff};\Delta^{eff})\right] ~ ,
\eea
which for discretized values of $t = an$ (with $n = 1, \dots N_T$) becomes
\bea
a_\mu^{\rm W}(t_0^{eff},t_1^{eff};\Delta^{eff}) & \equiv & 4 \alpha_{em}^2 \frac{1}{r_\mu^2}\frac{1}{(aX)^2}\sum_{n = 1}^{N_T} dt ~ K(r_\mu aXn) \, a^3 V^{ud}(an) \nonumber \\
& \cdot & \left[ \Theta(aXn,\tau_0;\tau_\Delta) - \Theta(aXn,\tau_1;\tau_\Delta)\right] ~ ,
\label{eq:amuWeff}
\eea
where
\bea
r_\mu & \equiv & m_\mu / X^{phys} ~ , \\
\tau_0 & \equiv & t_0 X^{phys} ~ , \\
\tau_1 & \equiv & t_1 X^{phys} ~ , \\
\tau_\Delta & \equiv & \Delta \, X^{phys} ~ ,
\eea
The main attractive feature of Eq.~(\ref{eq:amuWeff}) is that it involves both the vector correlator $a^3V^{ud}(an)$ and the quantity $(aX)$ both in lattice units.
Therefore, the knowledge of the lattice spacing is not required at all and the uncertainty of the scale setting cannot play any role.

An equivalent way to calculate $a^{\rm W}(t_0^{eff},t_1^{eff};\Delta^{eff})$ is to introduce the dimensionless variable
\be
\tau \equiv Xt ~ ,
\ee
which yields
\bea
a_\mu^{\rm W}(t_0^{eff},t_1^{eff};\Delta^{eff}) & \equiv & 4 \alpha_{em}^2 \frac{1}{r_\mu^2}\frac{1}{(aX)^3}\int_0^\infty d\tau ~ K(r_\mu \tau) \, a^3 V^{ud}(\tau / (aX)) \nonumber \\
& \cdot & \left[ \Theta(\tau,\tau_0;\tau_\Delta) - \Theta(\tau,\tau_1;\tau_\Delta)\right] ~ ,
\label{eq:amuWeffinterp}
\eea
where the vector correlator $a^3V^{ud}(\tau/(aX))$ can be obtained from the lattice data by smooth interpolation\footnote{We have explicitly checked that Eqs.~(\ref{eq:amuWeff}) and (\ref{eq:amuWeffinterp}) provide the same results (central values and errors).}.

\section{Intermediate-distance window}
\label{sec:winint}

\begin{figure}[htb!]
\centering
\includegraphics[scale=0.9]{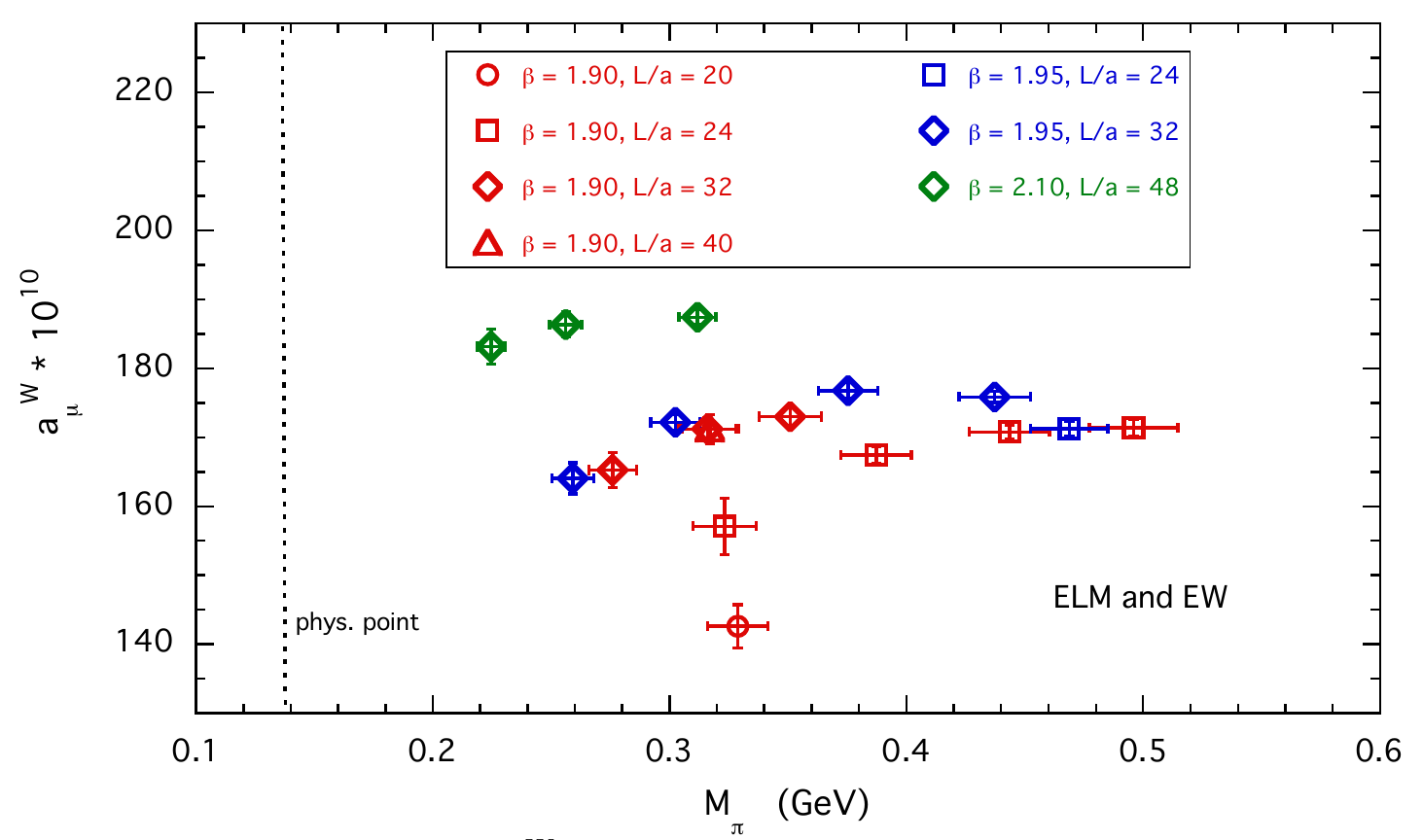}
\caption{Results obtained for $a_\mu^{\rm W} (ud)$ using Eq.~(\ref{eq:amuWeffinterp}) with $X = f_\pi$ and adopting the ETMC gauge ensembles employed in our study of the light-quark contribution to the HVP term in Ref.~\cite{Giusti:2018mdh}.}
\label{fig:amuW}
\end{figure}

Adopting the ETMC gauge ensembles of Ref.~\cite{Giusti:2018mdh} we try different choices of the hadronic quantity $X$ appearing in Eq.~(\ref{eq:amuWeff}), like the pion mass ($X = M_\pi$) or the pion decay
constant ($X = f_\pi$).
The goal is to achieve a dependence of $a_\mu^{\rm W}(ud)$ on the simulated pion mass as much flat as possible.
This can be obtained by using $X = f_\pi$ (with $f_\pi^{phys} = 130.4 ~ {\rm MeV}$) and the corresponding results for $a^{\rm W}_\mu (ud)$ are shown in Fig.~\ref{fig:amuW} versus the simulated pion mass $M_\pi$.
It can be seen that the dependence on $M_\pi$ is quite mild, while FVEs and discretization effects play a relevant role.

\begin{figure}[htb!]
\centering
\includegraphics[scale=1]{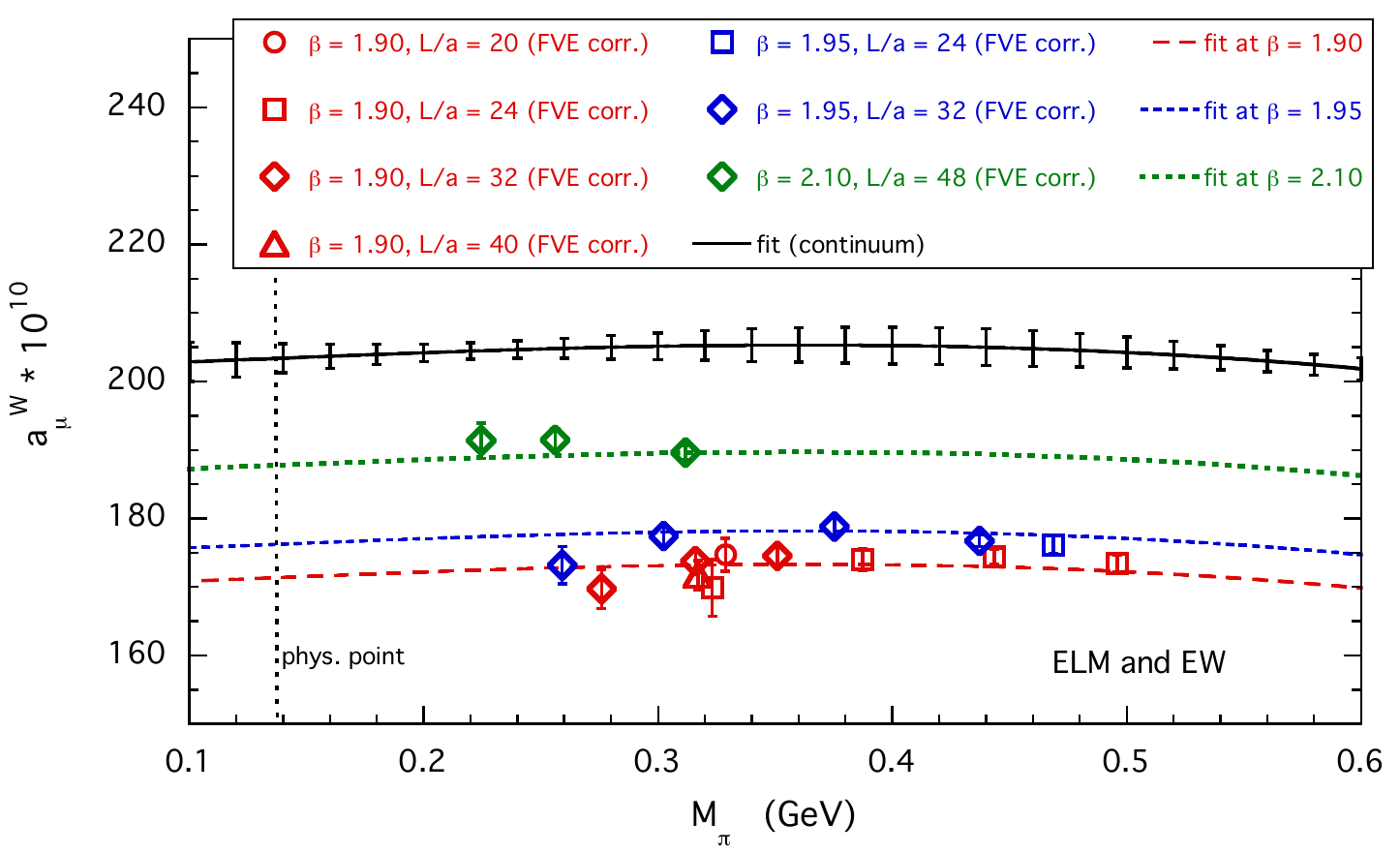}
\caption{Results obtained for $a_\mu^{\rm W}(ud)$ using a quadratic fit (i.e.~Eq.~(\ref{eq:amuWfit}) with $A_{1\ell} = 0$ and $A_2 \neq 0$). The lattice data are corrected for the FVEs determined in the fitting procedure. The dotted, short-dashed and dashed lines are the results of the fit (\ref{eq:amuWfit}) at fixed lattice spacing in the infinite volume limit. The solid line corresponds to the result (\ref{eq:amuWres}) in the continuum and infinite volume limit.}
\label{fig:amuWfit}
\end{figure}

We perform the extrapolations to the physical pion point ($M^{phys}_\pi = 135 ~ {\rm MeV}$) and to the continuum and infinite volume limits adopting the following phenomenological ansatz
\bea
a_\mu^{\rm W} (ud)& = & A_0 \left[ 1 + A_{1\ell} M^2_\pi \log{\left(M^2_\pi\right)} + A_1 M^2_\pi + A_2 M_\pi^4 + D_1 a^2\alpha_s^n(1/a) + D_2 a^4 \right] \nonumber \\
& \cdot & \left[ 1 + F M_\pi^2 e^{-M_\pi L} / (M_\pi L)^p \right] ~ ,
\label{eq:amuWfit}
\eea
where $A_0, A_{1\ell}, A_1, A_2, D_1, D_2$ and $F$ are free parameters.
We have tried several values of the power $p$, finding that the optimal choice is $p \simeq 2$.
Therefore, in what follows we put $p = 2$.
In Fig.~\ref{fig:amuWfit} the results obtained using a quadratic fit (i.e.~Eq.~(\ref{eq:amuWfit}) with $A_{1\ell} = 0$ and $A_2 \neq 0$) are shown against the lattice data corrected for the FVEs determined in the fitting procedure.

At the physical point our result is
\be
a^{\rm W}_\mu (ud)= 202.2 ~ (2.0)_{stat+fit} (0.4)_{chir} (1.5)_{disc} (0.7)_{FVE} [2.6] \cdot 10^{-10} ~ ,
\label{eq:amuWres}
\ee
where
\begin{itemize}
\item $()_{stat+fit}$ indicates the uncertainty induced by the statistical Monte Carlo errors of the simulations and its propagation in the fitting procedure;
\item $()_{chir}$ is the error due to the chiral extrapolation, estimated from the spread of the results corresponding to either a log ($A_{1\ell} \neq 0$ and $A_2 = 0$) or quadratic ($A_{1\ell} = 0$ and $A_2 \neq 0$) fitting function;
\item $()_{disc}$ is the uncertainty due to discretization effects. This is estimated in three ways: i) from the spread of the results corresponding to the use of the methods M1 and M2, which differ by ${\cal O}(a^2)$ effects, employed in Ref.~\cite{EuropeanTwistedMass:2014osg} to determine the relevant renormalization constants in the RI$^\prime$-MOM method; ii) by comparing the results corresponding to different choices of $n = 0, \dots, 3$ without fitting $a^4$ corrections (i.e.~$D_2 = 0$); iii) by fixing $n = 0$ and either including or excluding the $D_2 a^4$ term. We then combine those estimates according to Eq.~(28) of Ref.~\cite{EuropeanTwistedMass:2014osg};
\item $()_{FVE}$ is the uncertainty generated by FVEs, estimated by excluding the results corresponding to the two ensembles A40.XX with the smallest lattice size.
\end{itemize}

\begin{figure}[htb!]
\centering
\includegraphics[scale=0.22]{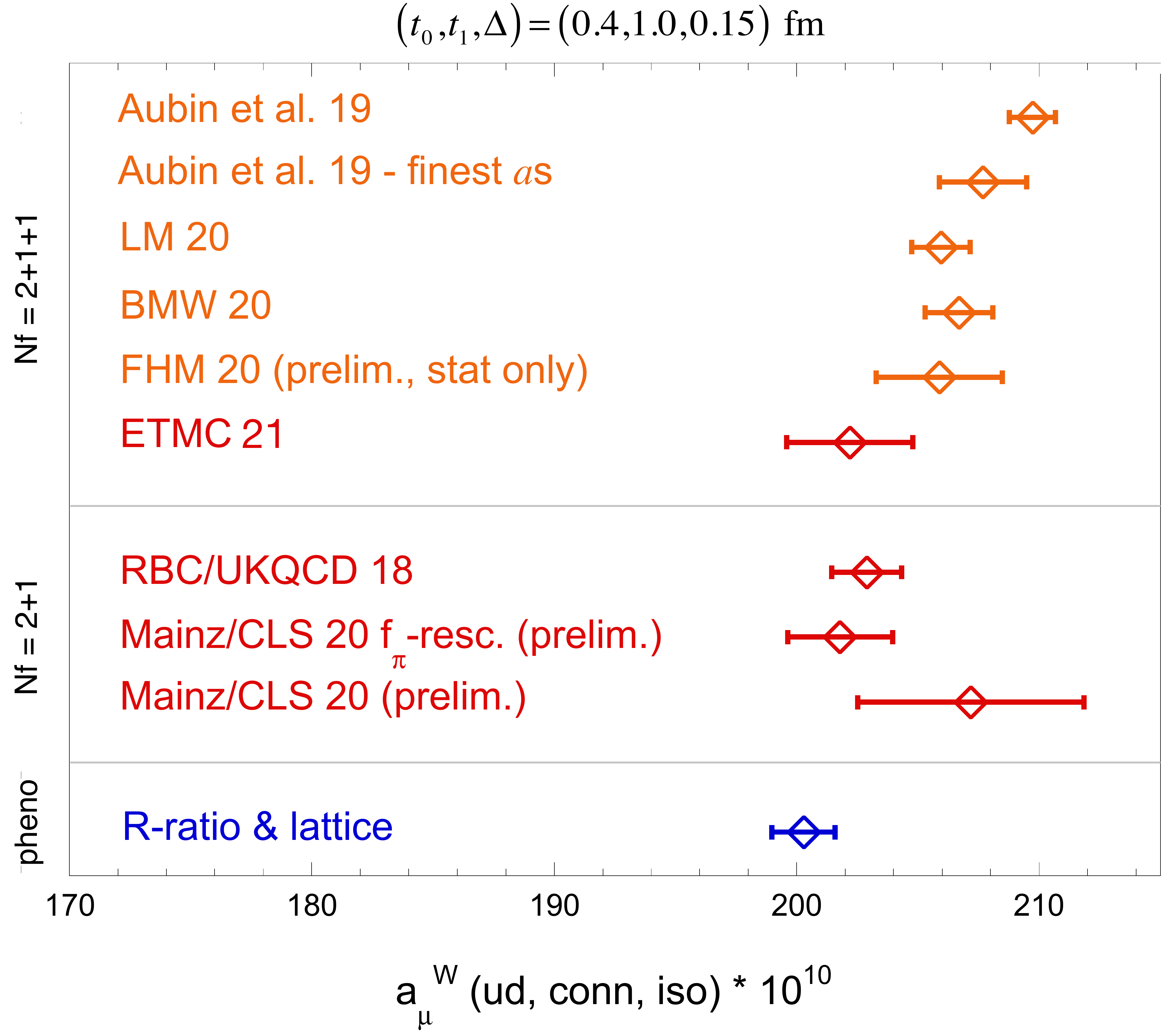}
\caption{Comparison of lattice and phenomenological results for the intermediate-distance window of the isospin-symmetric, light, quark-connected component of $a_\mu^{\rm HVP}$. Results correspond to the same choice of the parameters entering the window function (\ref{eq:winfunc}), namely $(t_0, t_1, \Delta) = (0.4, 1.0, 0.15)$ fm. Lattice determinations are classified according to the fermion action adopted ("Wilson-like" and staggered formulations indicated by red and orange data points, respectively) and the number of dynamical quarks used in the simulations. Each value is extracted by the following references: Aubin {\it et al.} \cite{Aubin:2019usy}, LM \cite{Lehner:2020crt}, BMW \cite{Borsanyi:2020mff}, FHM \cite{FHM:20}, RBC/UKQCD \cite{RBC:2018dos}, Mainz/CLS \cite{Mainz:20}, R-ratio \& lattice \cite{Borsanyi:2020mff}. The two determinations of Aubin {\it et al.} differ in the procedure used to perform the continuum extrapolation, while Mainz/CLS provides two preliminary estimates obtained by rescaling or not the raw lattice data.}
\label{fig:intermediate}
\end{figure}

So far, only four collaborations have published lattice results for $a_\mu^{\rm W}(ud)$, but some preliminary determinations have been recently presented.
In Fig.~\ref{fig:intermediate} we compare our finding (\ref{eq:amuWres}) both with other non-perturbative predictions and with a R-ratio estimate, obtained in \cite{Borsanyi:2020mff} by subtracting all lattice contributions, except the light-quark-connected one, from the phenomenological determination based on dispersive analyses of the experimentally measured $e^+e^- \to hadrons$ data.
Lattice results are classified according to the fermion action adopted and the number of dynamical quarks included in the simulations.
We observe some tensions between different estimates.
More importantly, a significant tension appears between the R-ratio prediction and the lattice determinations based on calculations with staggered sea quarks.

\section{Short- and long-distance windows}
\label{sec:winshortlong}

\begin{figure}[htb!]
\centering
\includegraphics[scale=1]{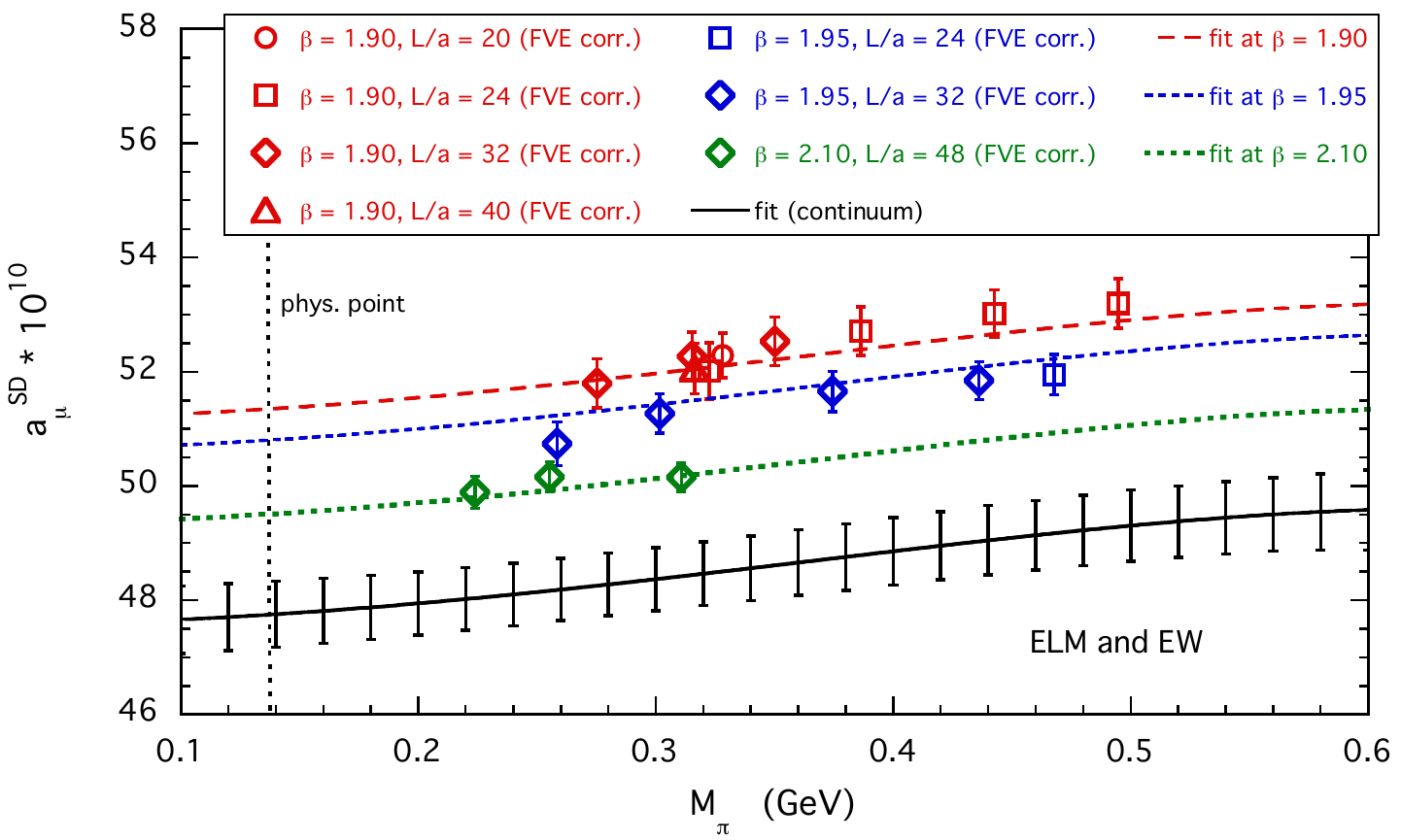}
\caption{The same as in Fig.~\ref{fig:amuWfit} but in the case of short-distance contribution $a_\mu^{\rm SD}$.}
\label{fig:amuSDfit}
\end{figure}

We repeat our ELM and EW procedures also in the case of the short-distance window contribution $a_\mu^{\rm SD}(ud)$.
The quality of the extrapolation to the physical pion point and to the continuum limit is illustrated in Fig.~\ref{fig:amuSDfit}.
The same fit function defined in (\ref{eq:amuWfit}) is adopted.
It can be seen that using $X = f_\pi$ the pion mass dependence is a bit more pronounced for $a^{\rm SD}_\mu(ud)$ with respect to $a_\mu^{\rm W}(ud)$.
At the physical point our result is
\be
a_\mu^{\rm SD} (ud)= 48.21 ~ (0.56)_{stat+fit} (0.10)_{chir} (0.50)_{disc} (0.25)_{FVE} [0.80] \cdot 10^{-10} ~ .
\label{eq:amuSDres}
\ee
where the error budget is estimated as in (\ref{eq:amuWres}). 

As far as the long-distance contribution $a_\mu^{\rm LD}(ud)$ is concerned, the use of ELM and EW procedures adopting either $X = M_\pi$ or $X = f_\pi$ does not lead to a mild pion mass dependence of $a_\mu^{\rm LD}(ud)$.
Moreover, the FVEs are large and they need to be treated in a non-perturbative way as done in Ref.~\cite{Giusti:2018mdh}.
Therefore, we limit ourselves to quote the result for $a_\mu^{\rm LD}(ud)$ obtained at the physical point using the analytic representation of the vector correlator $V^{ud}(t)$ of Ref.~\cite{Giusti:2018mdh}, namely
\be
a_\mu^{\rm LD} (ud)= 382.5 ~ (10.5)_{stat+fit} (5.2)_{syst} [11.7] \cdot 10^{-10} ~ .
\ee

Adopting the analytic representation of the vector correlator $V^{ud}(t)$ developed in Ref.~\cite{Giusti:2018mdh} for the SD and W windows too, we obtain
\bea
a_\mu^{\rm W} (ud)& = & 198.0 ~ (3.4)_{stat+fit} (4.7)_{syst} [5.8] \cdot 10^{-10} ~ , \\
a_\mu^{\rm SD} (ud)& = & 48.6 ~ (1.8)_{stat+fit} (1.0)_{syst} [2.0] \cdot 10^{-10} ~ ,
\eea
which in a reassuring way agree with the results (\ref{eq:amuWres}) and (\ref{eq:amuSDres}) within larger uncertainties.

\begin{figure}[htb!]
\centering
\includegraphics[scale=0.15]{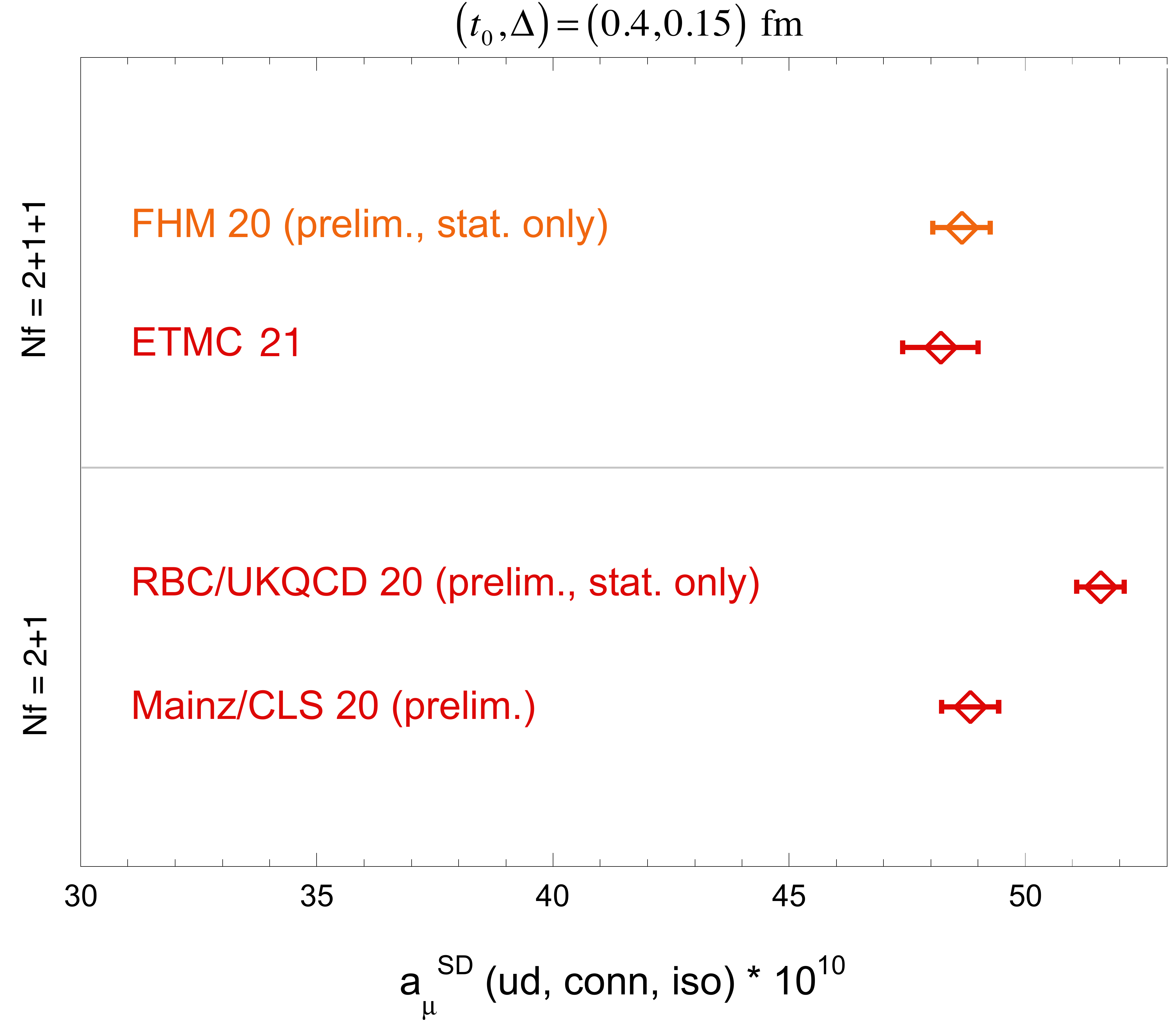}~\includegraphics[scale=0.15]{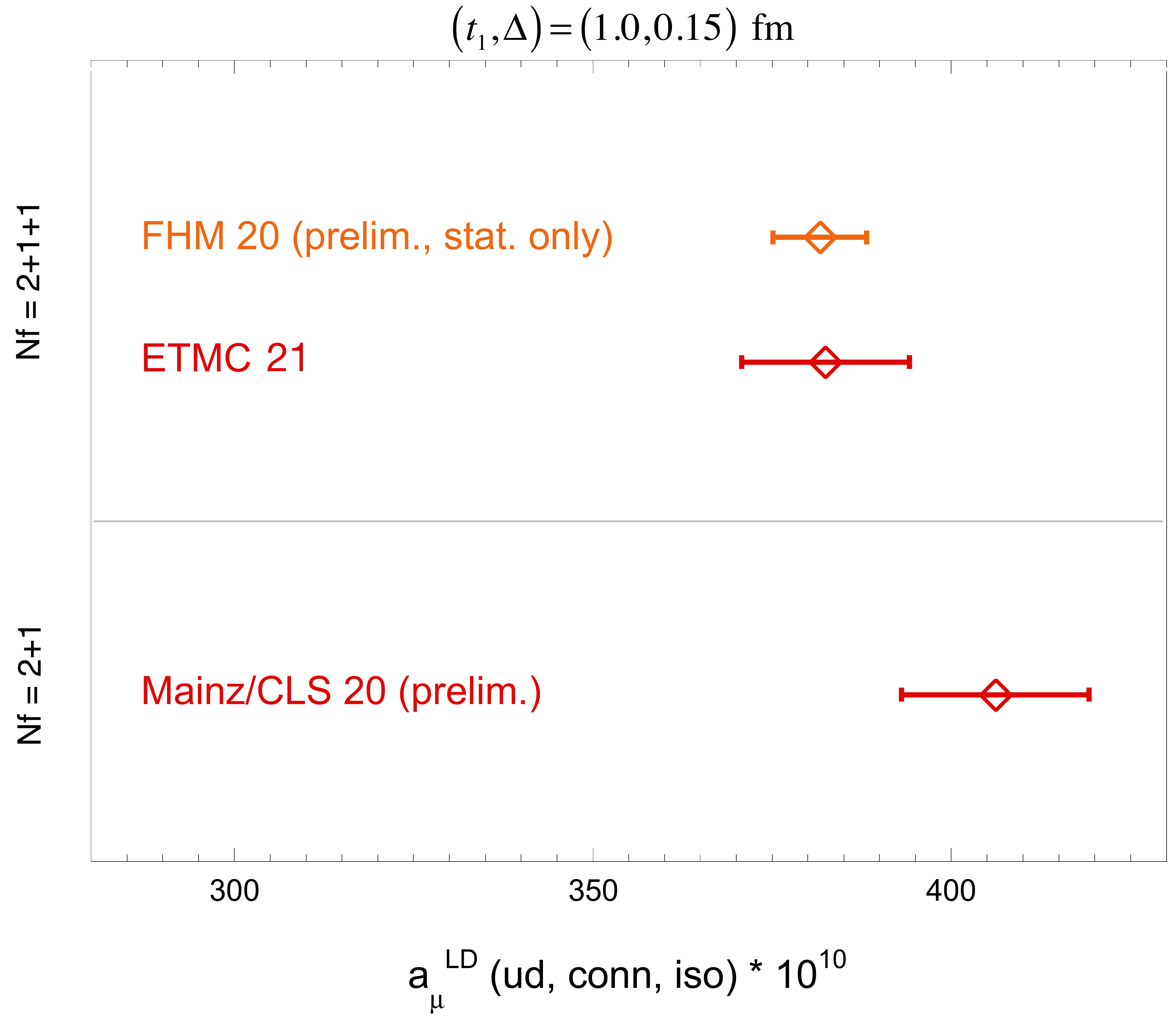}
\caption{The same as in Fig.~\ref{fig:intermediate} for the short- (left panel) and long- (right panel) distance windows $a_\mu^{\rm SD}(ud)$ and $a_\mu^{\rm LD}(ud)$.}
\label{fig:shortlong}
\end{figure}

As done for the intermediate-distance window in Sec.~\ref{sec:winshortlong}, in Fig.~\ref{fig:shortlong} we compare our determinations for $a_\mu^{\rm SD}(ud)$ and $a_\mu^{\rm LD}(ud)$ with lattice results available from other collaborations.

\section{Other contributions}
\label{sec:other}

We conclude the present contribution presenting some results for the intermediate-distance window of the strange, charm and isospin-breaking (IB) components of $a_\mu^{\rm HVP}$.
The results obtained at the physical point for each contribution are shown in Tab.~\ref{tab:otherres}.
The uncertainties represent the sum in quadrature of various sources of errors, namely statistical, fitting procedure, input parameters, discretization, FVEs and chiral extrapolation.

IB corrections contributing to orders ${\cal O}(\alpha_{em}^3)$ and ${\cal O}(\alpha_{em}^2(m_d - m_u)/\Lambda_{QCD})$ are calculated non-perturbatively within the RM123 approach \cite{deDivitiis:2013xla}, which consists in the expansion of the path integral in powers of the $u$- and $d$-quark mass difference $(m_d - m_u)$ and of the electromagnetic coupling $\alpha_{em}$.
The quenched-QED (qQED) approximation, which treats dynamical quarks as electrically neutral particles, is adopted and in Tab.~\ref{tab:otherres} an estimate of the error due to the qQED approximation is also included \cite{Giusti:2019xct}.

\begin{table}[htb!]
    \centering
      \begin{tabular}{cccc}
         \hline
         \hline
	 $f$ & $s$ & $c$ & IB \\
  	 \hline
          $a^{\rm W}_\mu (f) \cdot 10^{10}$ & $26.9 ~ (1.0)$ & $2.81 ~ (0.11)$ & $0.7 ~ (0.4)$ \\
          \hline
          \hline
	\end{tabular}
	\caption{Results for the intermediate-distance window of the strange, charm and IB quark-connected contributions to $a_\mu^{\rm HVP}$. The parameters entering the window function (\ref{eq:winfunc}) are set to $(t_0, t_1, \Delta) = (0.4, 1.0, 0.15)$ fm.}
	\label{tab:otherres}
\end{table}

\begin{figure}[htb!]
\centering
\includegraphics[scale=0.15]{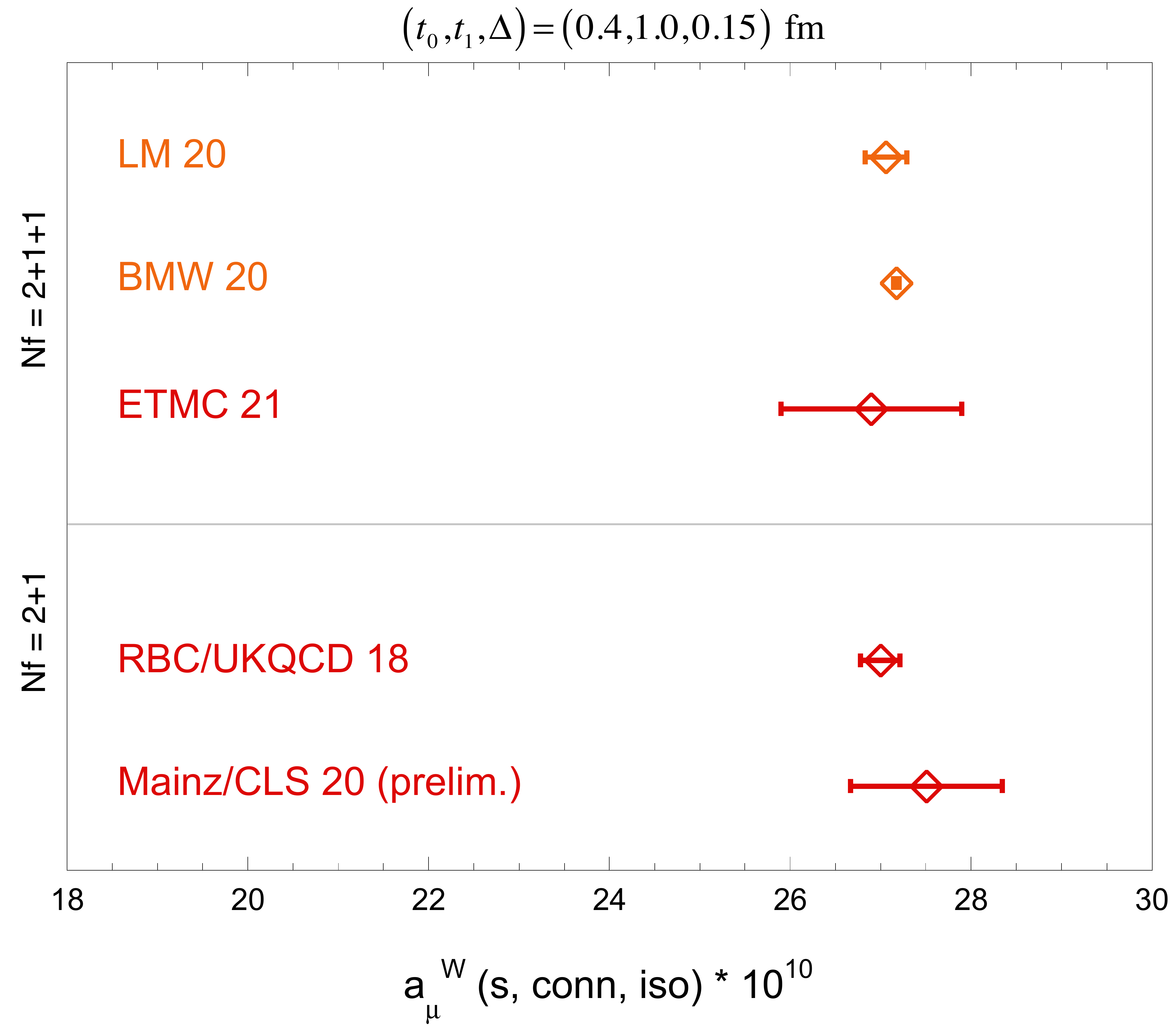}~\includegraphics[scale=0.15]{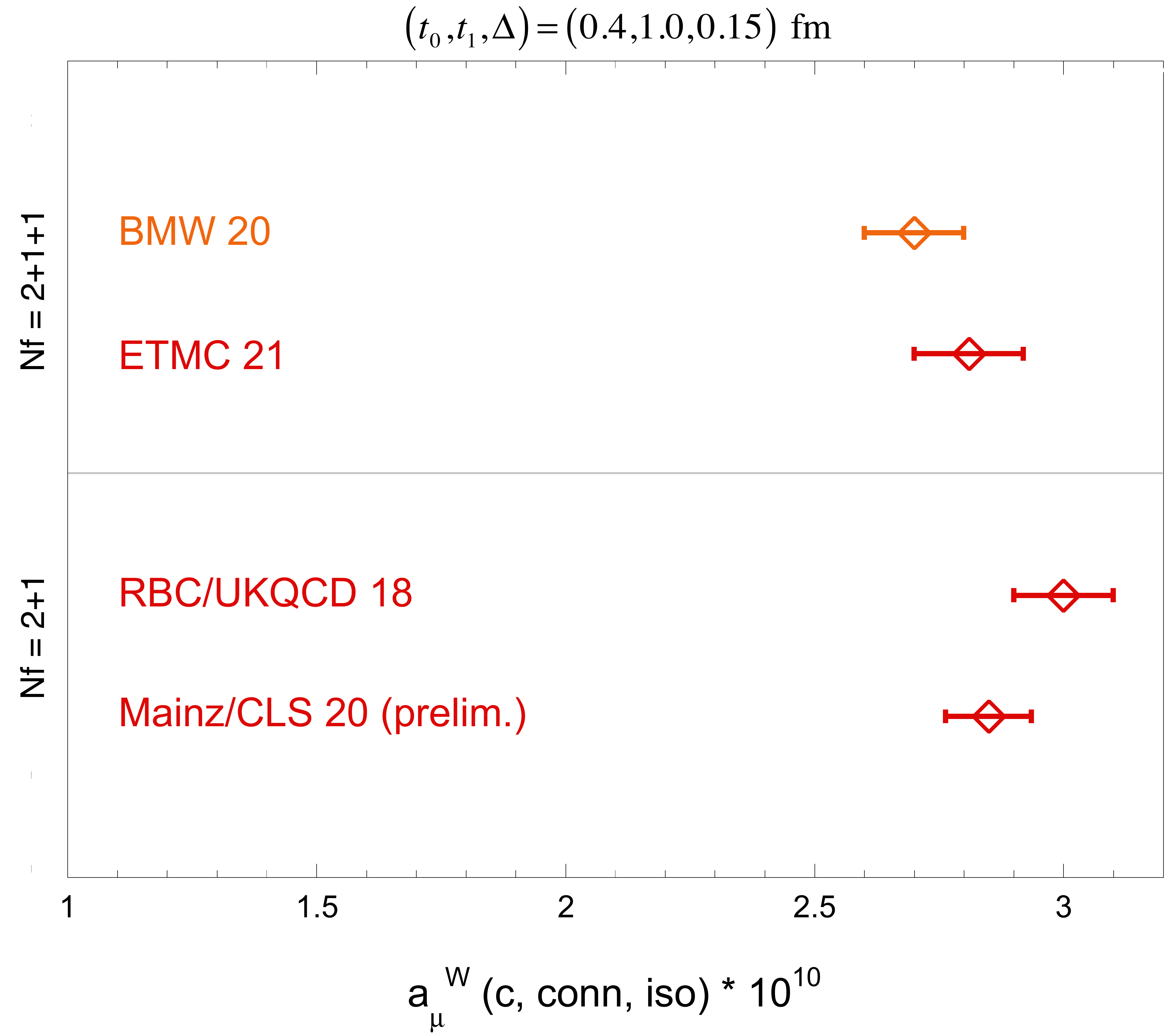}\\
\includegraphics[scale=0.15]{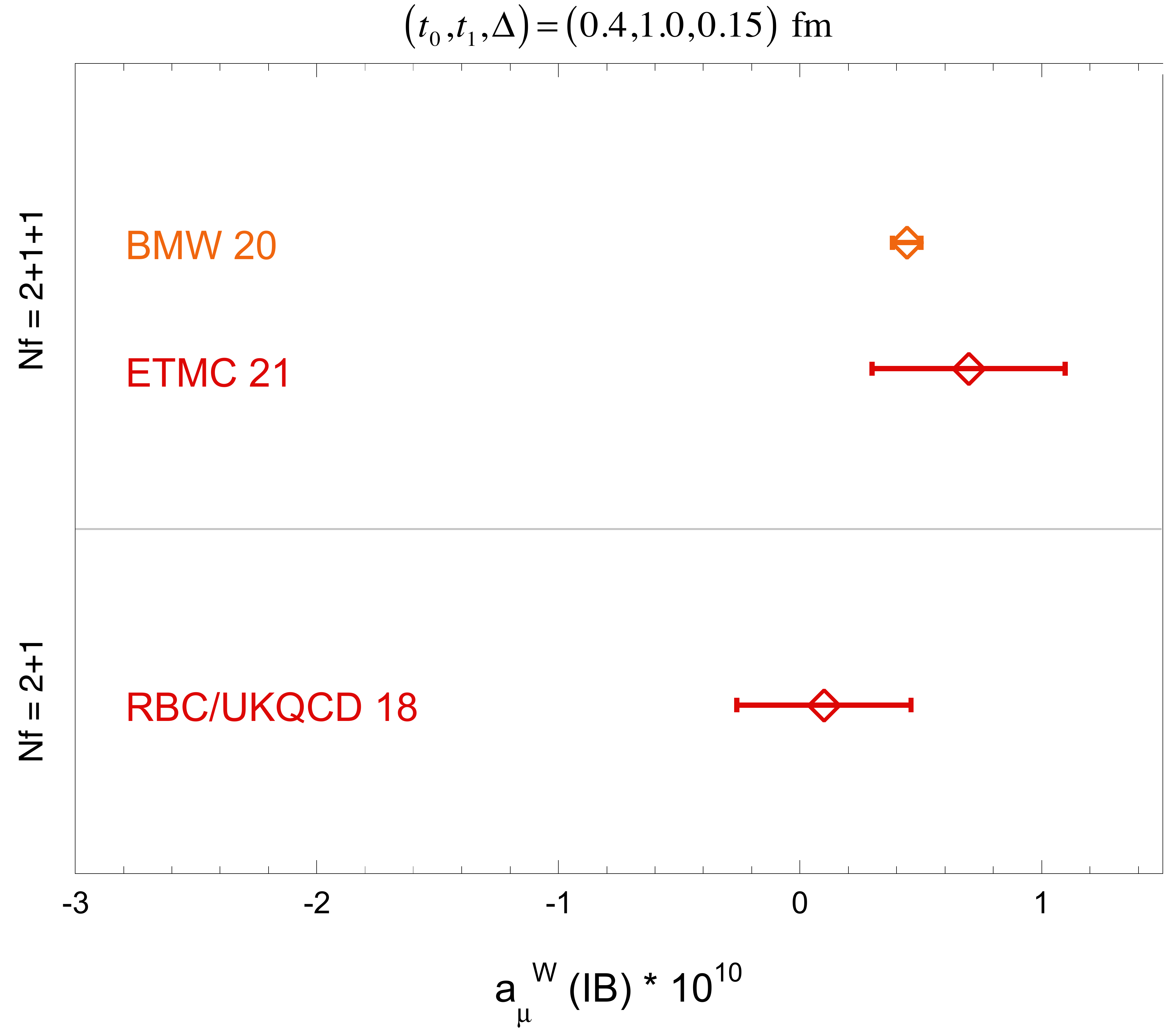}
\caption{The same as in Fig.~\ref{fig:intermediate} for $a_\mu^{\rm W}(s)$ (upper left panel), $a_\mu^{\rm W}(c)$ (upper right panel) and $a_\mu^{\rm W}({\rm IB})$ (lower panel).}
\label{fig:scIB}
\end{figure}

In Fig.~\ref{fig:scIB} other lattice results present in the literature are collected and compared with ours.

Using the findings of Refs.~\cite{Borsanyi:2020mff,RBC:2018dos} we estimate the contribution of the quark-disconnected diagrams to be equal to $a_\mu^{\rm W} (disconn.) = -0.9 ~ (0.2) \cdot 10^{-10}$.
Adding all the various contributions we get
\be
a_\mu^{\rm W} = 231.7 ~ (2.8) \cdot 10^{-10}\,,
\ee
which remarkably agrees well with the more precise R-ratio estimate $a_\mu^{\rm W} = 229.7 ~ (1.3) \cdot 10^{-10}$ from \cite{Borsanyi:2020mff}.

\end{document}